\documentstyle[12pt,aaspp4]{article}
\input psfig.sty
\def\be{\begin{equation}}
\def\ee{\end{equation}}
\def\ba{\begin{eqnarray}}
\def\ea{\end{eqnarray}}
\def\ref{\bibitem{}}
\def\~{$\sim$}

\lefthead{HICKSON}
\righthead{COMPACT GROUPS OF GALAXIES}

\begin{document}

\title{COMPACT GROUPS OF GALAXIES}

\author{\it Paul Hickson}

\affil{Department of Physics and Astronomy, University of British 
Columbia, 2219 Main Mall, Vancouver, B.C. V6T1Z4, Canada \\
paul@astro.ubc.ca}


\begin{abstract}

Compact groups of galaxies have posed a number of challenging questions.  
Intensive observational and theoretical studies are now providing answers to 
many of these, and at the same time, are revealing unexpected new clues about 
the nature and role of these systems.  Most compact groups contain a high 
fraction of galaxies having morphological or kinematical peculiarities, nuclear 
radio and infrared emission, and starburst or active galactic nuclei (AGN) 
activity.  They contain large quantities of diffuse gas and are dynamically 
dominated by dark matter.  They most likely form as subsystems within looser 
associations and evolve by gravitational processes.  Strong galaxy interactions 
result and merging is expected to lead to the ultimate demise of the group.  
Compact groups are surprisingly numerous, and may play a significant role in 
galaxy evolution.

\end{abstract}

\section{INTRODUCTION}

As their name suggests, compact groups are small systems of several galaxies in
a compact configuration on the sky.  The first example was found over one
hundred years ago by Stephan (1877) who observed it visually using the 40-cm 
refractor of the Observatoire de Marseille.  Stephan's Quintet is a small group
of five galaxies, three of which show strong tidal distortions due to 
gravitational interaction.  A second example was found 71 years later by 
Seyfert (1948) from a study of Harvard Schmidt plates.  Seyfert's Sextet 
(Figure 1) is one of the densest groups known, having a median projected 
galaxy separation of only $6.8 h^{-1}$ kpc (the Hubble Constant 
$H_0 = 100h$ km s$^{-1}$ Mpc$^{-1}$).

%--------------------------  figure 1
\begin{figure}
\centerline{\psfig{figure=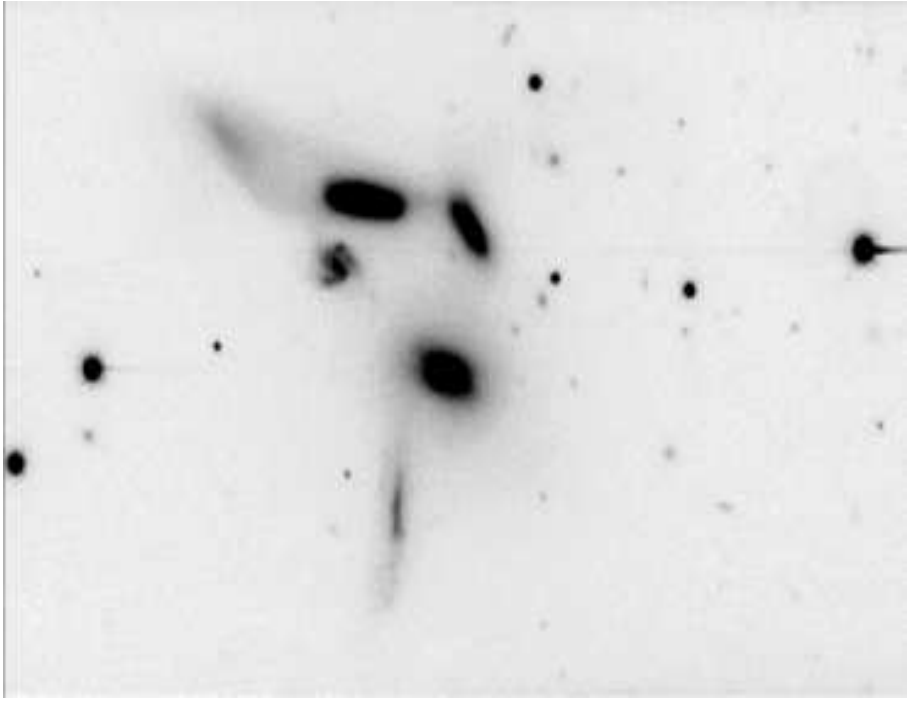,width=12cm,height=9cm}}
\vspace{1cm}
\caption{
Seyfert's Sextet. Discovered in 1948, this group of five galaxies is one of the
densest known. The sixth object appears to be a tidal plume. The small face-on 
spiral galaxy has a redshift that is more than four times larger than those of 
the other galaxies.
}
\end{figure}
%---------------------------------

The Palomar Observatory Sky Survey (POSS) provided a new and extensive resource
for the systematic investigation of small groups of galaxies.  Two catalogs, the
{\it Atlas of Interacting Galaxies} (Vorontsov-Velyaminov 1959, 1975) and the
{\it Atlas of Peculiar Galaxies} (Arp 1966), contain galaxies or galaxy groups
selected on the basis of visible signs of interaction or peculiar appearance.
In addition to Stephan's quintet and Seyfert's Sextet, these include many new
compact groups, including a striking chain of five galaxies, VV 172.  Prior to
these, Shakhbazian (1957) had discovered a small dense cluster of 12 faint red
galaxies that appeared so compact that they were initially mistaken for stars.
Over the next two decades, Shakhbazian and collaborators examined over 200 POSS
prints covering 18\% of the sky and cataloged 376 additional ``compact groups of
compact galaxies'' (Shakhbazian 1973, Shakhbazian \& Petrosian 1974, Baier
et al 1974, Petrosian 1974, 1978, Baier \& Tiersch 1975-79).  Apart from
occasional photographic or spectroscopic observations (eg.  Mirzoian et al
1975, Tiersch 1976, Massey 1977, Shakhbazian \& Amirkhanian 1979,
Vorontsov-Velyaminov et al 1980, Vorontsov-Velyaminov \& Metlov 1980) these
systems initially received little attention.  However, interest in them is
growing.  Although the majority seem to be small clusters, they share some of
the properties, and pose some of the same questions, as do compact groups.

When redshifts were measured for galaxies in the first compact groups (Burbidge
\& Burbidge 1959, 1961a), surprises were found.  Both Stephan's Quintet and
Seyfert's Sextet contain a galaxy with a discordant redshift.  It seemed
unlikely that a foreground or background galaxy would appear so often projected
within such compact systems (Burbidge \& Burbidge 1961a).  This impression was
further reinforced with the discovery of yet another discordant redshift in
VV 172 (Sargent 1968).  Are these examples of physical association between
objects of widely different redshifts, as has been advocated for many years by
Arp (1987)?

Even if the discordant galaxies are ignored, the velocity dispersions of these
systems are generally higher than would be expected given their visible mass
(Burbidge \& Burbidge 1959, 1961b, 1961c, Burbidge \& Sargent 1971).  It was
argued that such groups must be unbound and disrupting (eg.  Ambartsumian 1961),
although Limber and Mathews (1960) showed that the virial theorem could be
satisfied for Stephan's Quintet, given the uncertainties in the projection
factors, if the individual galaxy masses were considerably larger than those of 
isolated galaxies. The observations can of course also be explained if the bulk 
of the mass is in a non-visible form.  In hindsight, this was one of the 
earliest indications of the possible existence of dark matter in galactic 
systems.

A new problem emerged with the realization that bound groups would be
unstable to orbital decay resulting from gravitational relaxation processes 
(Peebles 1971).  A simple calculation indicated that the dynamical-friction 
timescale was much shorter than the Hubble time (Hickson et al 1977), as was 
soon confirmed by numerical simulations (Carnevali et al 1981).  At the same 
time, it became increasingly clear that mergers played an important role in the 
evolution of many, if not all, galaxies (Press \& Schechter 1974, Ostriker \& 
Tremaine 1975). Compact groups emerged as prime locations for investigations of 
the dynamical evolution of galaxies.

Motivated by the desire for a homogeneous sample that could be subject to
statistical analysis, Rose (1977), and later, Hickson (1982) produced the first
catalogs of compact groups having specific, quantitative, selection criteria.
Subsequent detailed investigation, at many wavelengths, has produced a large
body of observational data for the Hickson catalog.  As a result, it has now
become possible to address some of the outstanding questions concerning the
nature of compact groups and their role in galaxy evolution.  Not surprisingly,
new questions have been raised and new controversies have appeared.  However,
much progress has been made in resolving both old and new issues.

This review is organized as follows:  In Section 2, the definition of a compact
group is discussed, along with methods of identification and surveys that have
been made.  In Sections 3-5, observed properties of these systems are summarized
and discussed.  Sections 6 and 7 focus mainly on interpretation of the
observations, and on implications of these results.  All work on compact groups 
of galaxies cannot possibly be discussed in this short paper, although an
attempt is made to touch upon most current topics.  Other recent reviews of
compact groups and closely related subjects include those by White (1990),
Hickson (1990, 1997), Whitmore (1992), Kiseleva \& Orlov (1993), Sulentic
(1993), and Mamon (1995).

\section{IDENTIFICATION AND SURVEYS}

By ``compact group'', we mean a small, relatively isolated, system of typically
four or five galaxies in close proximity to one another.  Such groups do not
necessarily form a distinct class, but may instead be extreme examples
of systems having a range of galaxy density and population.  Because
of this, the properties of the groups in any particular sample may be strongly
influenced by the criteria used to define the sample.  The early surveys used
qualitative criteria that, while successful in finding many interesting
individual objects, do not easily allow one to draw broad conclusions about the
groups as a whole.  Thus, the focus in recent years has been on samples selected
using specific, quantitative, criteria.  These criteria define the
minimum number and magnitude range of the galaxies, and also consider the galaxy
spatial distribution.

The use of quantitative selection criteria was pioneered by Rose (1977) who
searched for groups that have three or more galaxies that are brighter than a 
limiting magnitude of 17.5, and that have a projected surface density 
enhancement of a factor of 1000 compared to the surrounding background galaxy 
density.  Searching an area of 7.5\% of the sky, he found 170 triplets,
33 quartets, and 2 quintets. Unfortunately, the sample received little 
follow-up study.  Sulentic (1983) re-examined the 35 Rose groups which contain 
four or more galaxies and found that only a third actually satisfied the 
selection criteria.  This is testimony to the difficulty of visual searches.  
A more fundamental problem is that the fixed magnitude limit in the selection 
criteria makes the sample susceptible to strong distance-dependent biases.

In an attempt to reduce such effects, Hickson (1982, 1993, 1994) adopted a
relative magnitude criterion, selecting systems of four or more galaxies whose
magnitudes differ by less than 3.0.  A distance independent (to first order) 
compactness criterion was employed:  $\bar\mu_G < 26$, where $\bar\mu_G$ is the
mean surface brightness of the group calculated by distributing the flux of the
member galaxies over the smallest circular area containing their geometric
centers.  To avoid including the cores of rich clusters, an isolation criterion
was necessary so as to reject the group if a non-member galaxy, not more than 
3 mag fainter than the brightest member, occurred within three radii of the 
center of the circle. (A non-member galaxy is a galaxy which if included in the
group would cause the group to fail one or more of the selection criteria.) 
From a search of 67\% of the sky (all the POSS prints), and using magnitudes 
estimated from the POSS red prints, exactly 100 groups were found satisfying 
these criteria (hereafter HCG's). As the HCG sample is now the most widely 
studied, it is important to examine the biases introduced by the criteria, and 
by the visual search procedure.  

Any sample selected on the basis of surface density will suffer from geometric 
and kinematic biases.  The former occurs because non-spherical systems will be
preferentially selected if they are oriented to present a smaller
cross-sectional area (eg.  prolate systems pointed towards us); the latter
because we will preferentially select systems that, owing to galaxy orbital
motion, are momentarily in a more compact state (transient compact
configurations).  Thus, a compact group might result from a chance alignment or
transient configuration within a loose group (Mamon 1986).  This question will
be considered in more detail in Section 6.2.

Other biases arise from the subjective nature of the search procedure.  The
original catalog contains a few mis-identifications, such as compact galaxies
mistaken for stars, and marginal violations of the isolation criteria.  In
addition, when photometry was obtained for the galaxies in the catalog (Hickson
et al 1989), it was found that some groups would not satisfy the selection 
criteria if photometric magnitudes are used.  Attempts to rectify these 
problems have been made by Hickson et al (1989) and Sulentic (1997).  However, 
it should be emphasized that changes, such as using photometric magnitudes in 
the selection criteria, are not corrections to the catalog, but are actually 
the imposition of {\it additional} a postiori selection criteria.  The 
resulting subsample is by no means complete because the new criteria are 
applied only to the visually-selected HCG catalog and not to the entire sky.

Because of the difficulty of identifying faint groups, the HCG catalog starts to
become significantly incomplete at an integrated magnitude of about 13 (Hickson
et al 1989, Sulentic \& Raba\c{c}a 1994).  A more subtle effect results from the
difficulty of recognizing low-surface-brightness groups.  Because of this, the
catalog also becomes incomplete at surface brightnesses fainter than 24 (Hickson
1982).  Yet another effect is that groups may be more noticeable if the magnitude
spread of their members is small.  Thus, the catalog may also be incomplete for
magnitude intervals greater than about 1.5 (Prandoni et al 1994).  These
effects are of critical importance in statistical analyses of the sample.  One
immediate conclusion is that the actual number of groups which satisfy the
selection criteria may be considerably larger than the number found by a
subjective search.

It has recently become feasible to find compact groups by automated techniques.
Mamon (1989) used a computer to search Tully's (1987) catalog of nearby galaxies
and identified one new compact group, satisfying Hickson's criteria, in the 
Virgo Cluster.  Prandoni et al (1994) applied similar criteria to digital scans 
of $\sim 1300$ deg$^2$ around the southern galactic pole and detected 59 new 
southern compact groups (SCGs).  Observations are presently underway to obtain 
accurate photometry and redshifts for this sample (Iovino private 
communication). The digitized Palomar Sky Survey II also offers new 
opportunities for the identification of compact groups (De Carvalho \& 
Djorgovski 1995).

An alternative approach is to identify groups of galaxies from redshift
information, as was first done by Humason et al (1956). With the advent of 
large-scale redshift surveys, it has become possible to identify a 
reasonably-large sample of compact groups in this way. Barton et al (1996) have 
compiled a catalog of 89 redshift-selected compact groups (RSCGs) found by 
means of a friends-of-friends algorithm applied to a complete magnitude-limited 
redshift survey.  Galaxies having projected separations of $50h^{-1}$ kpc or 
less and line-of-sight velocity differences of 1000 km s$^{-1}$ or less are 
connected and the sets of connected galaxies constitute the groups.  The 
numerical values were chosen to best match the characteristics of the HCG 
sample, and indeed, many of those RSCGs that have at least four members are 
also HCGs.  There are some significant differences, however: Because foreground 
and background galaxies are automatically eliminated by the velocity selection 
criteria, this technique is more effective at finding groups in regions of 
higher galaxy density, which would fail the HCG isolation criterion.  This
criterion requires that the distance to the nearest neighbor be at least as
large as the diameter of the group.  The RSCG criteria, on the other hand, 
require only that the nearest-neighbor distance be larger than the threshold
distance ($50h^{-1}$ kpc for the RSCGs), which may be considerably smaller.  
It will therefore allow the inclusion of groups that are physically less 
isolated (and therefore less physically distinct) than would the HCG criterion. 
One would also expect that the numbers of groups found in a given volume by the 
less-restrictive RSCG criteria to be larger than by the HCG criteria, as seems 
to be the case.  While the redshift-selection method compliments the HCG 
angular-selection technique, one should keep in mind that it also selects 
groups according to apparent (projected) density -- the velocity information 
serves only to reject interlopers.  Thus it will be subject to some of the 
biases discussed above.  Also, because the galaxy sample used is 
magnitude-limited, rather than volume-limited, there will be redshift-dependent 
biases in the RSCGs.  However, the well-defined selection criteria and the
completeness of the sample, should allow a quantitative determination of the
effects of this bias.

\section{SPACE DISTRIBUTION AND ENVIRONMENT}

The space distribution and environment of compact groups provides important
clues to their nature.  The median redshift of the HCGs is $z = 0.030$, placing
most of them at distances well beyond the Virgo Cluster (Hickson et al 1992).
A cursory inspection reveals that they are fairly uniformly distributed and show
no preference for rich clusters.  This is at least partly due to the isolation
criterion.  However, galaxies in rich clusters have rather different kinematical
and morphological properties than do those in compact groups, so one might
justifiably argue that small clumps of galaxies within clusters are not compact
groups.

A natural question is whether or not compact groups are associated with loose
groups.  Rood \& Struble (1994) observed that 70\% of the HCGs are located
within the bounds of cataloged loose groups and clusters.  Studies of the
distribution of galaxies in redshift space (Vennik et al 1993, Ramella et al
1994, Sakai et al 1994, Garcia 1995, Barton et al 1996) indicate that compact
sub-condensations do occur within loose groups and filaments.  Vennik et al 
(1993) and Ramella et al (1994) find that most HCGs are indeed associated with 
loose groups.

While the above studies show that compact groups trace large-scale structure, it
is also clear that they prefer low-density environments.  Sulentic (1987), Rood
\& Williams (1989), Kindl (1990), and Palumbo et al (1995) have examined the
surface density of galaxies surrounding the groups.  They generally agree that
about two-thirds groups show no statistically significant excess of nearby
neighbors.  This is not inconsistent with the redshift-space results because 
most of the HCG associations identified by Ramella et al (1996) contain fewer 
than 5 excess galaxies.  Thus while compact groups are associated with loose 
groups and filaments, these tend to be low-density and sparsely-populated 
systems.

Are the galaxies in compact groups in any way distinct from those in their
immediate environments?  Rood \& Williams (1989) and Kindl (1990) both found
that compact groups, including those in rich environments, contain a
significantly smaller fraction of late-type (spiral and irregular) galaxies than
do their neighborhoods.  This result is of particular importance to the
question of the physical nature of compact groups, discussed below in Section 6.  
In addition, many independent studies have found significant differences 
between galaxies in compact groups and those in other environments.  These are 
examined in Section 5.

\section{DYNAMICAL PROPERTIES}

The first studies of individual compact groups (Burbidge \& Burbidge 1959,
1961b, 1961c, Burbidge \& Sargent 1971, Rose \& Graham 1979, Kirshner \& 
Malumuth 1980) indicated short dynamical times and mass-to-light ratios 
intermediate between those of galaxies and rich clusters. However, because of 
the small number of galaxies, estimates of the space velocities and physical 
separations of galaxies in individual groups are highly uncertain. Meaningful 
dynamical conclusions about systems containing only four or five galaxies 
requires statistical analysis of large homogeneous samples.

By 1992, velocities had been measured for almost all 462 galaxies in the HCG
catalog (Hickson et al 1992).  The distribution of galaxy velocities, relative
to the median of each group, is shown in Figure 2.  It can be seen that
the majority of velocities fall within a roughly Gaussian distribution of
standard deviation $\sim250$ km s$^{-1}$ (the sharp peak at zero velocity
results from the use of the median).  This characteristic velocity is quite
similar to velocity dispersions found in loose groups, and much smaller than
typical velocity dispersions in rich clusters.  In addition to the Gaussian
core, a flat component is seen in the velocity distribution of Figure 2.
This is expected as some galaxies which are not physically related to the group
will appear projected on the group by chance.  This component contains about
25\% of the total number of galaxies.  Whether of not chance projection can
account for such a large number of ``discordant'' galaxies is still a matter of
some debate, and is discussed further in Section 6.1.

%--------------------------  figure 2
\begin{figure*}
\centerline{\psfig{figure=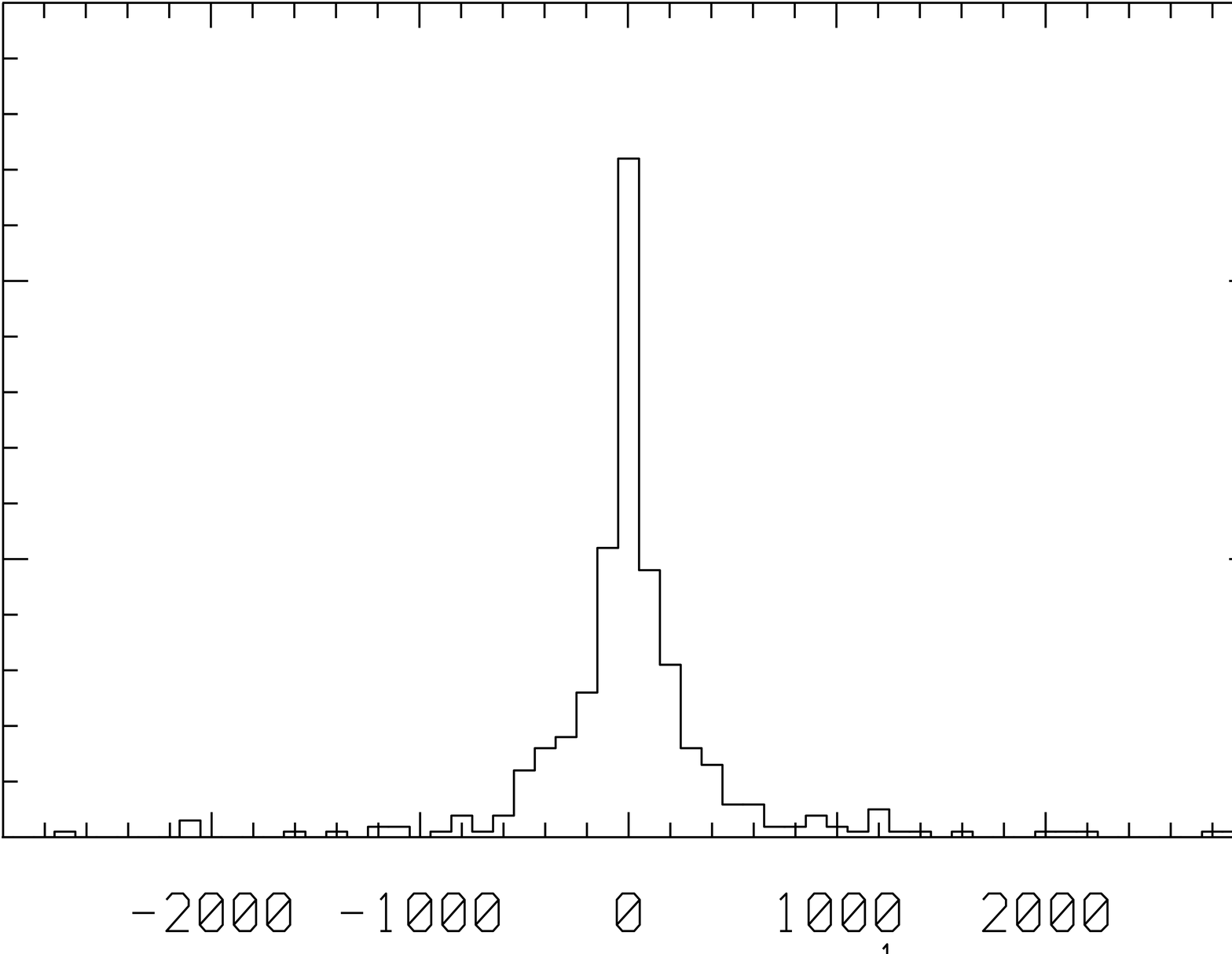,width=12cm,height=9cm}}
\vspace{1cm}
\caption{
Velocity distribution of galaxies in compact groups. The figure shows
the distribution of the difference between the observed galaxy radial 
velocity and the median velocity of galaxies in the group to which
it belongs, for 410 galaxies in the HCG catalog. Most galaxies (77\%)
have velocity differences less than 500 km s$^{-1}$ from the median.
}
\end{figure*}
%---------------------------------

For a system of characteristic linear size $R$ and internal velocity $V$, a
characteristic dynamical time is $t_d = R/V$.  A characteristic mass density is
$\rho = 1/Gt_d^2$, from which one can estimate the total mass within the region 
occupied by the galaxies.  After removing galaxies whose velocities differ from 
the group median by more that 1000 km s$^{-1}$, Hickson et al (1992) found that 
$t_d \sim 0.02 H_0^{-1}$ and obtained a mean mass-to-light ratio of $50h$ 
(solar units) for the HCGs.  Similar values were found for several Shakhbazian 
groups (Tikhonov 1986, Amirkhanian \& Egikian 1987, Amirkhanian 1989, Lynds et 
al 1990, Amirkhanian et al 1991). Since the mass-to-light ratios of individual 
HCG galaxies is on the order of $7h$ (Rubin et al 1991), the galaxies appear to 
contain only $\sim 15$\% of the total mass.

\section{STRUCTURE AND MORPHOLOGY}

The spatial distribution and luminosities of the member galaxies provides 
further clues to the nature of compact groups.  If they are primarily 
projections or transient configurations, the luminosity function should be the 
same as that of the parent systems, and the spatial distribution of the 
galaxies should be consistent with a random distribution.  If they are bound 
physical systems, the luminosity function and spatial distribution might show 
features that reflect the origin or subsequent evolution of compact groups.

\subsection{Shapes and orientations}

The shapes of compact groups were first investigated by Arp (1973) who concluded
that galaxy ``chains'' were unusually predominant.  However, Rose (1977)
determined that the ellipticities of his groups were consistent with a random
distribution of galaxies.  Using the larger HCG sample, Hickson et al (1984)
and Malykh \& Orlov (1986) reached the same conclusion as Arp -- the groups are
typically more elongated than would be a random distribution of galaxies.  An
immediate consequence of this result is that compact groups cannot easily be
explained as random projections or chance crossings, as this would largely erase
any inherent ellipticity of a parent loose group.  From static simulations,
Hickson et al (1984) concluded that the observed ellipticities are best matched
by three-dimensional shapes that are intrinsically prolate.  The same result was
found by Oleak et al (1995) in a recent study of the shapes of 95 Shakhbazian
compact groups.  These conclusions, however, are not unique.  Hickson et al
(1984) also found the shapes to be consistent with those seen in dynamical
simulations of compact groups seen in projection as subgroups within loose
groups.  In addition, one must always be concerned about possible selection
biases.  It may be that highly elongated groups (such as VV 172) are more easily
noticed in visual searches.  It will be interesting to see if these results are
confirmed by studies of groups found by automated searches.

If the intrinsic shapes of compact groups are related to their formation
process, one might expect to see a relationship between the orientation angle of
a group and the environment.  Palumbo et al (1993) examined the environments of
the HCGs, and found that the orientations of the major axes of the groups were
consistent with an isotropic distribution.

If compact groups are not simply projection effects, they might be expected to
show a centrally-concentrated surface density profile, as is seen in clusters of
galaxies.  Although the number of galaxies in individual compact groups is
small, with a large sample it is possible to estimate a mean profile.  By
scaling and superimposing the HCGs, Hickson et al (1984) found evidence for
central concentration.  Mendes de Oliveira \& Girard (1994), using a similar
analysis, concluded that the mean surface density profile is consistent with a
King (1962) model with typical core radius of 15$h^{-1}$ kpc.  Most recently,
Montoya et al (1996) have analyzed the profiles of the 42 HCG quartets which
have accordant redshifts.  Their technique uses the distribution of projected
pair separations and thus avoids assumptions about the location of the group
center.  They find a smaller core radius ($6h^{-1}$ kpc for a King model).  The
fact that Montoya et al (1996) find a consistent density profile for all
groups, without any scaling, is particularly interesting.  This would not be
expected if most groups are chance alignments within loose groups.  It also 
implies that compact groups have a unique scale, which seems counter to the 
concept of hierarchical clustering.  Montoya et al (1996) suggest that this 
arises as a result of a minimum mass density and velocity dispersion that is 
required for the groups to be virialized (Mamon 1994).

\subsection{Compact-group galaxies}

There have been several studies of the morphological types of galaxies in
compact groups (Hickson 1982, Williams \& Rood 1987, Sulentic 1987, Hickson
et al 1988b).  Most studies agree that the fraction $f_s$ of late
type galaxies is significantly less in compact groups than in the field. 
Hickson et al (1988b) find $f_s = 0.49$ for the HCGs; Prandoni et al (1994) 
obtain $f_s = 0.59$ for the SCGs.  Both these values are substantially lower 
than those found for field galaxy samples ($f_s \simeq 0.82$, Gisler 1980, 
Nilson 1973).

Also well established is morphological type concordance, observed in both the
HCGs and SCGs (Sulentic 1987, Hickson at al 1988b, Prandoni et al 1994).  A 
given compact group is more likely to contain galaxies of a similar type 
(early or late) than would be expected for a random distribution.  White 
(1990) has pointed out that such concordance could result from a correlation 
of morphological type with some other property of the group.  The strongest 
such correlation found to date is between morphological type and velocity 
dispersion (Hickson et al 1988b). As Figure 3 shows, groups with higher
velocity dispersions contain fewer late-type (gas-rich) galaxies.  They also
tend to be more luminous.  The importance of velocity dispersion, in addition 
to local density, on the galaxy morphology had previously been emphasized by 
De Souza et al (1982).  A crucial clue is that the morphology-density relation 
seen in clusters and loose groups (Dressler 1980, Postman \& Geller 1984,
Whitmore \& Gilmore 1992) is not the dominant correlation in compact groups 
(Hickson et al 1988b), although some effect is present (Mamon 1986). This 
suggests that the velocity dispersion is more fundamental, ie of greater 
physical relevance to the formation and evolution of galaxies in compact 
groups, than is apparent physical density.

%--------------------------  figure 3
\begin{figure*}
\centerline{\psfig{figure=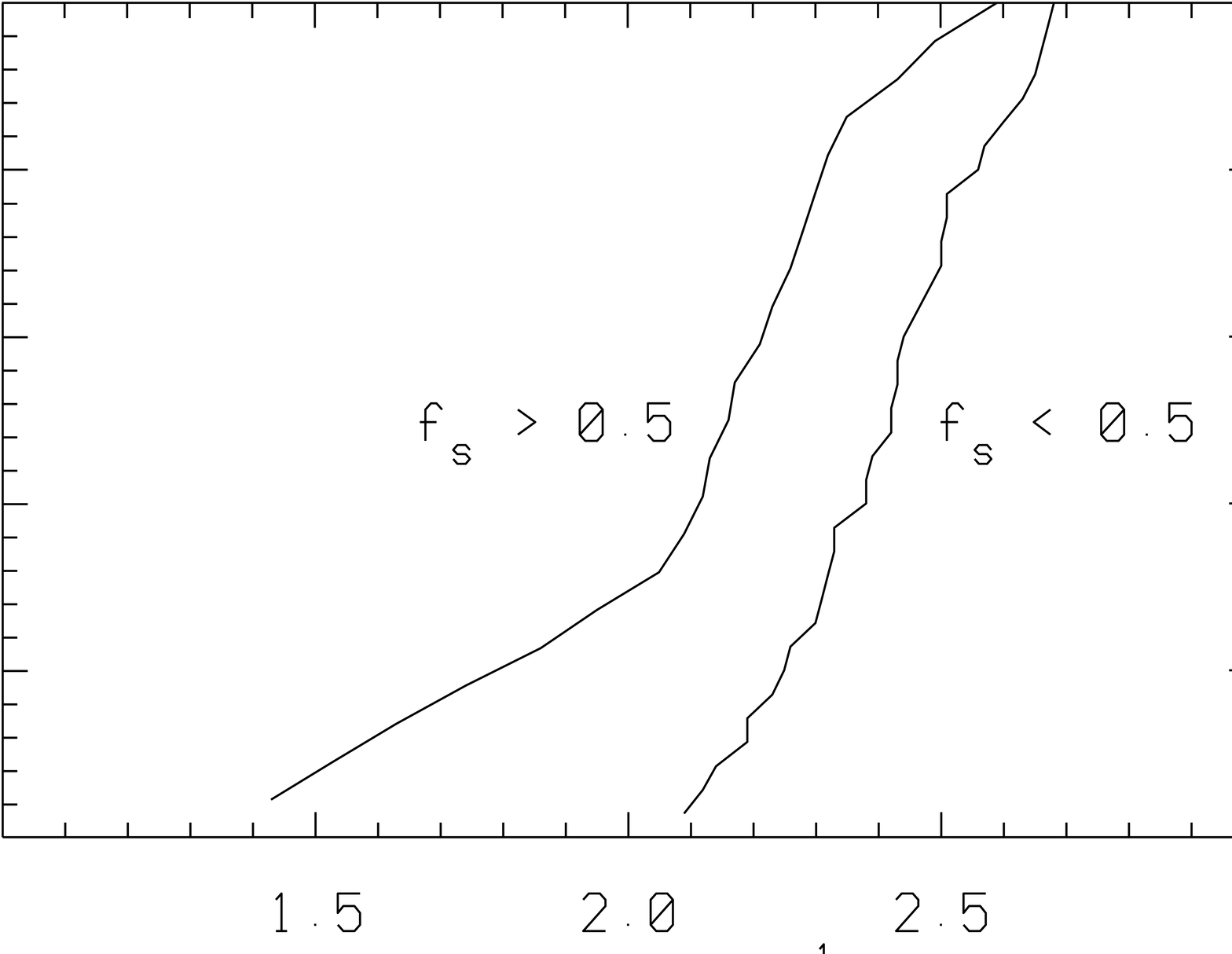,width=12cm,height=9cm}}
\vspace{1cm}
\caption{
Morphology-velocity correlation for compact groups. The figure shows the
cumulative distributions of velocity dispersion for spiral rich
($f_s > 0.5$) and spiral-poor ($f_s < 0.5$) groups. The former have
typically half the velocity dispersion than the latter and a broader
velocity range.
}
\end{figure*}
%---------------------------------

There is much evidence that interaction is occurring in a large fraction of
galaxies in compact groups.  The strongest direct support comes from kinematical
studies.  Rubin et al (1991) found that two thirds of the 32 HCG spiral 
galaxies that they observed have peculiar rotation curves.  These show 
asymmetry, irregularity and in some cases extreme distortion, characteristic of 
strong gravitational interaction.  This result has recently been challenged by 
Mendes de Oliveira et al (preprint) who obtained H$_\alpha$ velocity maps for 
26 HCG spiral galaxies and found that only 1/3 showed abnormal rotation curves. 
They suggest that the difference is due to the more-complete spatial sampling 
of their data.

In their study, Rubin et al (1991) observed 12 HCG elliptical galaxies and
detected nuclear emission in 11 of them.  This high fraction suggests that
interactions and mergers may be supplying gas to these galaxies.  This idea
received independent support from radio observations in which neutral hydrogen
emission was detected in three compact groups which contain only elliptical
galaxies (Huchtmeier 1994).

Zepf \& Whitmore (1993) found that elliptical galaxies in compact groups tend to
have lower internal velocity dispersions than do ellipticals in other 
environments having the same effective radii, absolute magnitudes and colors.  
They therefore do not lie on the fundamental plane defined by other elliptical 
galaxies.  This discrepancy correlates with isophote shape in that those 
galaxies that have ``disky'' or irregular isophotes tend to have lower 
velocity dispersion.  Both Zepf \& Whitmore (1993) and Bettoni \& Fasano (1993,
1995, 1996, Fasano \& Bettoni 1994) report that HCG elliptical galaxies are 
less likely to have ``boxy'' isophotes, and more likely to have irregular 
isophotes.  Such effects are consistent with results of simulations of tidal 
encounters (Balcells \& Quinn 1990).

\subsection{Optical luminosity function}

The luminosity function of compact groups was first estimated by Heiligman
\& Turner (1980).  They examined a sample consisting of Stephan's Quintet,
Seyfert's Sextet and eight more compact groups from the Arp and 
Vorontsov-Velyaminov catalogs, and concluded that compact groups contain 
relatively fewer faint galaxies than does a comparable field galaxy sample.  
Analysis of the relative luminosities within individual HCGs (Hickson 1982), 
and studies of several Shakhbazian groups (Kodaira et al 1991), showed a 
similar effect, although Tikhonov (1987) found a luminosity function similar 
to that of field and cluster galaxies.

The larger HCG sample allows the question of the galaxy content of compact
groups to be addressed with greater certainty.  The standard technique for
determination of the luminosity function weights each galaxy by $V_m/V$, where
$V$ is the volume of the smallest geocentric sphere containing the group, and
$V_m$ is the volume of the largest such sphere within which the group could have
been detected.  Using this approach Sulentic and Raba\c{c}a (1994) obtained a
luminosity function for HCG galaxies similar to that of field galaxies.
However, Mendes de Oliveira and Hickson (1991) argued that the standard
calculation does not address the selection effects of the HCG sample.  For
example, the luminosity range within an individual group is limited by the
3-mag range of the selection criteria.  Because of this, fainter
galaxies within compact groups are not included in the catalog.  In order to
account for such biases they used a modeling technique in which galaxies were
drawn from a trial luminosity function and assigned to groups.  Redshifts were
given to each group according to the observed distribution, and groups that
failed to meet the HCG selection criteria were rejected.  The luminosity
distribution of the resulting galaxy sample was then compared to the
observations and the process repeated with different trial luminosity functions.
Their best-fit luminosity function is deficient in faint galaxies, although a
normal field-galaxy luminosity function is not excluded.

To avoid the selection problem, Ribeiro et al (1994) obtained deeper
photometry for a subsample of the HCGs in order to include the fainter galaxies
explicitly.  Since redshifts are not known for these galaxies, a correction for
background contamination was made statistically.  The luminosity function that
they obtained is similar to that of field galaxies.  Most recently, the
luminosity function for the RSCGs has been computed by Barton et al (1996).
They find it to be mildly inconsistent with that of field-galaxies, in the same
sense as that of Mendes de Oliveira \& Hickson (1991) for the HCGs.  Figure 4 summarizes these estimates of the
luminosity function of compact galaxies in terms of the Schechter (1976)
parameters $M^*$ and $\alpha$.

%--------------------------  figure 4
\begin{figure*}
\centerline{\psfig{figure=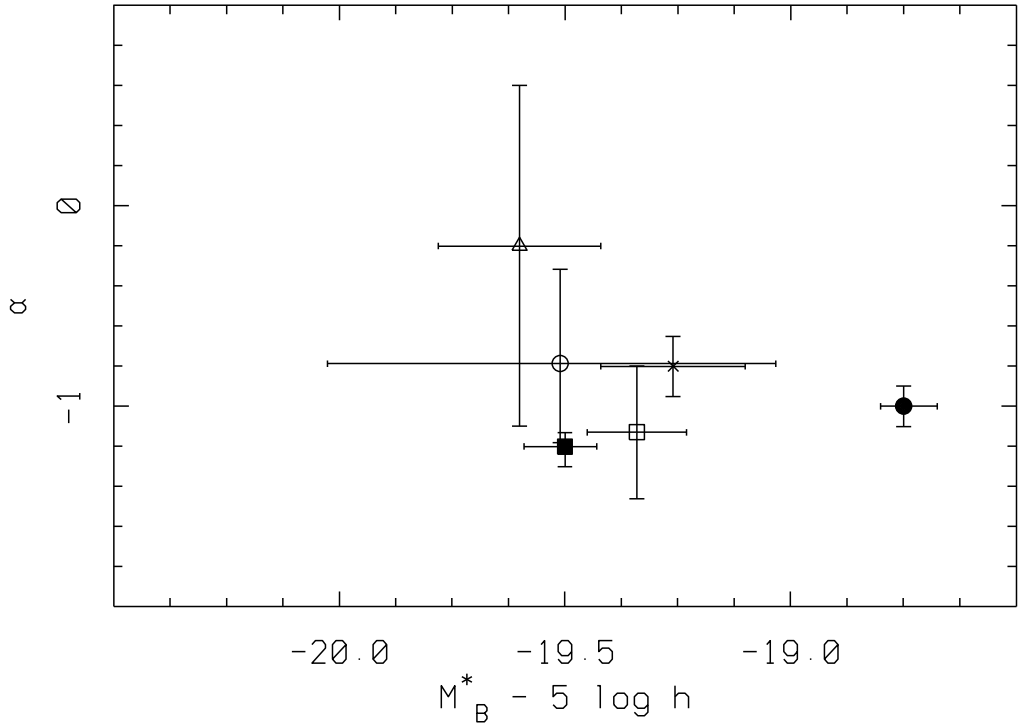,width=12cm,height=9cm}}
\vspace{1cm}
\caption{
Optical luminosity function parameters of compact groups. Triangle:
Mendes de Oliveira \& Hickson (1991), open square: Sulentic \& Raba\c{c}a
(1994), cross: Ribeirao et al (1994), open circle: mean of the
three RSCG samples (Barton et al 1996). For comparison, the filled
circle and square indicate luminosity function parameters for galaxies in
the CFA-Combined (Marzke et al 1994) and SSRS2 (da Costa et al 1994) 
surveys respectively, representing galaxies in lower-density environments.
}
\end{figure*}
%---------------------------------

How do we interpret these apparently conflicting results?  Prandoni et al
(1994) have argued that the HCG catalog is biased toward groups with a small
magnitude range $\Delta m$, because the SCGs have a larger fraction of high
$\Delta m$ groups.  However, it is not known what fraction of such groups are
physically real, as few redshifts have yet been obtained.  Such a bias could
affect the luminosity function of Mendes de Oliveira \& Hickson (1991), 
particularly at the faint end, but it is not  evident that the bias is 
sufficient to account for the apparent faint-galaxy deficit. On the other hand, 
the small sample used by Ribeiro et al (1994) may not be representative of 
compact groups in general. Hickson (1997) points out that the Ribeiro et al
(1994) sample has a spiral fraction of 0.60, substantially higher than that of 
the whole HCG catalogue, and contains 7 of the 16 HCGs found to be in a high 
density environment by Palumbo et al (1995). This suggests that the their 
sample has more than the usual amount of field galaxy contamination. De 
Carvalho et al (1994), note that the faint galaxies form a more extended 
distribution than do the brighter galaxies. Thus they may be a dynamically 
distinct component, or simply unrelated field galaxies. Finally, the 
redshift-selected RSCG sample also shows mild evidence for a faint galaxy 
deficiency.

While the faint end of the LF in compact groups appears to be 
depleted, there is evidence that the bright end may be enhanced. Limber and 
Matthews (1960) were the first to remark that ``the members of Stephan's
Quintet are to be classed among the brightest of galaxies''. It is possible 
that this may be in part due to interaction-induced star formation, at least 
for the spiral galaxies. On the other hand Mendes de Oliveira \& Hickson (1991) 
compared their luminosity function of elliptical galaxies in compact groups 
with those in the Virgo and Coma Clusters (as reported by Sandage et al 1985 
and Thompson \& Gregory 1980) and found that that the compact-group ellipticals 
have a luminosity enhancement of more than 1 mag compared to cluster 
ellipticals. Sulentic and Raba\c{c}a (1994) find a similar enhancement in 
their morphological-type-specific luminosity function. This suggests that 
compact group elliptical galaxies may have a unique formation mechanism.

From the luminosity function, one can estimate the contribution of
compact groups of galaxies ${\cal L}_{CG}$ to the total galaxian luminosity 
density ${\cal L}$. Mendes de Oliveira \& Hickson (1991) obtained a ratio of 
${\cal L}_{CG}/{\cal L} \simeq 0.8\%$. Applying the same analysis to the 
luminosity function of Ribeiro et al (1994) gives a ratio of 3.3\%. For the 
RSCGs, the figure is comparable: for groups of four or more galaxies, Barton 
et al (1996) obtain a compact group abundance of $1.4 \times 10^{-4} h^{-3}$ 
Mpc$^{-1}$ which leads to a luminosity density ratio of approximately 3\%. 
These are surprisingly high figures considering the short dynamical times of 
most compact groups.

\subsection{Star formation and nuclear activity}

Evidence is accumulating that tidal interactions play an important role in 
triggering starburst activity in galaxies (eg. Maccagni et al 1990,
Campos-Aguilar \& Moles 1991, Kormendy \& Sanders 1992, Sanders \& Mirabel 
1996). Compact groups, with their high galaxy density and evident signs of 
galaxy interaction should be ideal systems in which to study such effects. 
Many HCGs do in fact contain galaxies showing starbursts or harboring active 
galactic nuclei (AGN). For example, HCG 16 is found to contain a Seyfert 2 
galaxy, two LINERs, and three starburst galaxies (Ribeiro et al 1996). HCG 31 
contains five galaxies showing signs of recent starburst activity (Rubin et al 
1990, Iglesias-P\'aramo \& V\'ilchez preprint). Seyfert galaxies are also found 
in HCG 77, 92, 93 and 96.

The general degree of star formation activity in compact group galaxies can
be determined from infrared observations.  To date, studies have been based 
primarily on data from the IRAS satellite. Hickson et al (1989) found sources 
in 40 HCG from a search of the Point Source Catalog. They concluded that the 
ratio of far-infrared-to-optical luminosity is greater by about a factor of 
two in compact group galaxies, compared to that of isolated galaxies. This 
result was disputed by Sulentic and De Mello Raba\c{c}a (1993) who argued that 
the low spatial resolution of the data made the assignment of infrared flux to 
individual galaxies ambiguous. They concluded that redistribution of the flux 
could result in little or no infrared enhancement, a conclusion echoed by 
Venugopal (1995). However, in cases of doubt, Hickson et al (1989) identified 
the infrared galaxy on the basis of radio emission. The well-known correlation
between infrared and radio continuum emission makes it unlikely that the 
results are much in error. Analysis of improved data (eg. Allam et al 1996) 
should soon resolve questions about the identifications and infrared fluxes.

The resolution problem can be avoided by considering the infrared colors of the 
sources instead of the infrared/optical ratio. Zepf (1993) compared the ratio 
of 60 $\mu$m to 100 $\mu$m fluxes of compact group galaxies with those of 
isolated galaxies and also with those of galaxies believed to be currently 
merging. He found that the compact group sample was significantly different 
from both other samples, and estimated that about 1/3 of the compact group 
galaxies had warm colors (larger 60/100 $\mu$m ratios) similar to those of 
merging galaxies.

Another approach to interpreting the infrared results was taken by Menon (1991) 
who emphasized that the strong correlation between radio and infrared radiation 
indicates that these likely originate from a common region. In compact group 
spirals the radio emission is primarily nuclear whereas in isolated spirals it 
originates in the disk. If this is also true for the infrared flux, there must 
be an enhancement of the infrared/optical ratio, in the nuclear region, of more 
than an order of magnitude. This idea is supported by recent 
millimeter-wavelength observations (Menon et al 1996) in which CO emission is 
detected in 55 of 70 IRAS-selected HCG galaxies. The inferred ratio of 
infrared luminosity-to-H$_2$ mass showed an enhancement which correlates with 
the projected nearest-neighbor distance. 

Further clues are provided by radio continuum studies. Nonthermal emission
from spiral galaxies can arise from both disk and nuclear sources. Disk
emission is predominantly due to supernova remnants and is thus related
to the star formation rate. Nuclear emission can arise both from star
formation and from an active nucleus. Menon (1995) observed 133 spiral 
galaxies in 68 HCG, and found that overall they typically show less continuum 
emission than those in isolated environments, which is consistent with the 
neutral hydrogen observations. However, when considering the nuclear regions 
alone, the radio emission is found to be an order of magnitude higher 
compared to isolated spirals. The implication is that star formation and/or 
AGN activity is substantially enhanced in the nuclear regions of many compact 
group spiral galaxies. This is generally consistent with a picture in which 
galaxy interactions remove gas from the outer regions of galaxies, while 
simultaneously allowing gas to flow inwards toward the nucleus, resulting in 
enhanced star formation in the nuclear region, and possibly fueling an active 
nucleus.

Although there is a clear example of tidal interaction stimulating disk radio
emission in at least one compact group (Menon 1995a), statistical evidence for 
a link between interactions and radio emission in compact groups is only now 
accumulating. If interactions are stimulating nuclear radio emission, one 
would expect the radio luminosity to be correlated with some index describing 
the degree of interaction such as the projected distance to the nearest 
neighbor. Evidence in support of this was found by Vettolani \& Gregorini 
(1988) who observed that early-type galaxies which have a high ratio of 
radio-to-optical emission show an excess of nearby neighbors. A similar effect 
was observed by Malumian (1996) for spiral galaxies in groups. Examining 
compact group galaxies, Menon (1992) found that elliptical and S0 galaxies 
detected at a wavelength of 20 cm had closer neighbors than the undetected 
galaxies. The effect was not found for spiral galaxies, but if one considers
only the detected galaxies, there is a significant correlation between
radio-to-optical luminosity and nearest neighbor distance for both early and 
late type galaxies (TK Menon, private communication).

Continuum radio emission has also been detected in a number of HCG elliptical 
galaxies. Unlike those found in cluster ellipticals, the radio sources are 
low-luminosity and compact.  Where spectral indices are available, they 
indicate that the radio emission arises from an AGN rather than from starburst 
activity (TK Menon, private communication). In the HCG sample, there is a 
significant preference for radio-loud elliptical galaxies to be first-ranked 
optically (Menon \& Hickson 1985, Menon 1992). The probability of radio 
emission does not correlate with absolute luminosity, but instead correlates
with {\it relative} luminosity within a group. Spiral HCG galaxies do not show 
this effect. Although the tendency of radio galaxies in rich clusters to be 
first-ranked has been known for many years, it is surprising to find a
similar effect in small groups, where the number of galaxies and luminosity 
range is small, the gravitational potential well is much less clearly 
defined, and it is unlikely that any individual galaxy holds a central 
location. The effect of optical rank on radio emission had been previously 
noted in other small groups by Tovmasian et al (1980) although these authors 
made no distinction between elliptical and spiral galaxies. It is difficult 
to imagine any explanation for this result in which the compact group is not a 
true physical system. It would appear that, regardless of absolute luminosity, 
only the first-ranked (presumably the most massive in the group) elliptical 
galaxy can develop a radio source. 

\subsection{Diffuse light}

Stars stripped from galaxies by tidal forces should accumulate in the potential 
well of the group. and may be detectable as diffuse light. In an early 
photographic study, Rose (1979) found no evidence for diffuse light in his 
groups, and was lead to the conclusion that most of his groups must be
transient configurations. However, Bergvall et al (1981) were successful in 
detecting ionized gas and a common halo around a compact quartet of interacting 
early-type galaxies, and evidence for a common halo in VV 172 was reported by 
Sulentic \& Lorre (1983). Diffuse light can clearly be seen in HCG 94, and has 
been found in HCG 55 (Sulentic 1987), but Pildis et al (1995b) did not detect 
any in seven other compact groups. Analysis by Mamon (1986) indicated that 
while the expected diffuse light should be detectable with modern techniques, 
it would generally be very faint. Estimates of the total amount of diffuse 
light in the detected groups are rather uncertain as they depend sensitively on 
subtraction of the galactic light, and the sky background. Deeper photometry 
and improved image processing techniques may yet reveal diffuse light in other 
compact groups (Sulentic 1997).

\subsection{Cool gas}

The mass and distribution of cool galactic and intergalactic gas, can be
obtained from observations of the 21-cm line of neutral hydrogen. The first 
such study of a large sample of compact groups is that of Williams \& Rood 
(1987) who found a median HI mass of $2.2 \times 10^{10} M_\odot$. They 
concluded that compact groups are typically deficient in neutral hydrogen by 
about a factor of two compared to loose groups. This effect is consistent with 
similar deficit in continuum radio emission seen in the disks of compact group 
spiral galaxies (Menon 1995),  and suggests that interactions in compact groups 
has removed much of the gas from the galaxies. Simulations suggest that in 
addition to an outflow of gas, inflow also occurs which may fuel nuclear 
star-formation, as suggested by the strongly-enhanced radio emission seen in 
the nuclear regions of compact group spiral galaxies (Menon 1995). 

High-resolution studies of individual groups (Williams \& van Gorkom 1988, 
Williams et al 1991) showed clearly that the gas is not confined to the 
galaxies. In two of three groups studied, the radio emission originates from a 
common envelope surrounding the group and in the third group there are signs of 
tidal distortion. These results strongly indicate that at least these compact 
groups are physically dense systems and not chance alignments or transient 
configurations in loose groups. They also show that many groups have evolved to 
the point that gas contained within individual galaxies has been distributed 
throughout the group.

In contrast to the HI results, Initial CO-line observations of 15 compact-group 
galaxies (Boselli et al 1996) indicated a normalized molecular gas content
similar to that of isolated spiral galaxies. However, this result is based on 
normalizing the flux by the optical area of the galaxy, rather than by the 
infrared luminosity, and may be biased by the relatively small sizes of
compact-group galaxies. Further CO studies, currently in progress, should soon 
settle this question.

\subsection{Hot gas}

X-ray observations of hot gas in clusters of galaxies can reveal the
amount, distribution, temperature and metallicity of the gas, as well as the 
relative amount and distribution of the total gravitating mass. Temperature,
metallicity (fraction of solar abundance), and bolometric luminosities
are estimated by fitting a spectral model, such as that of Raymond \& Smith
(1977) to the data. X-ray emission from compact ``poor clusters'' was first 
reported by Schwartz et al (1980), who concluded that their X-ray properties 
were similar to those of rich clusters. Using the Einstein observatory, 
Bahcall et al (1984) first detected X-ray emission from Stephan's quintet. The 
X-ray map revealed that the emission is diffuse and not centered on individual 
galaxies. However, we now know that most of this emission is associated
with a shock front rather than gas trapped in the group potential well
(Sulentic et al 1995). Although several other groups (Bahcall et al 1984, 
Biermann \& Kronberg 1984) were detected by the Einstein observatory, further 
progress required the improved sensitivity of the ROSAT X-ray observatory.

Pointed ROSAT observations revealed massive hydrogen envelopes surrounding the 
NGC 2300 group, a bright elliptical-spiral pair with two fainter members 
(Mulchaey et al 1993), and HCG 62, a compact quartet of early-type galaxies 
(Ponman \& Bertram 1993), and showed that these systems are dominated by dark 
matter. Subsequent investigations detected X-rays from 18 additional compact 
groups, either from individual galaxies, or from diffuse gas (Ebeling et al 
1994, Pildis et al 1995a, Sarraco \& Ciliegi 1995, Sulentic et al 1995). 
These studies showed that the physical properties of individual systems span a 
wide range, but that the ratio of gas-to-stellar mass is significantly lower 
than in rich clusters. Moreover, the detected compact groups all contained a
majority of early-type galaxies. No spiral-rich groups were detected (although 
Mulchaey et al 1996b pointed out that they might be found from QSO absorption 
spectra). This result is consistent with the fact that X-ray-selected 
groups (Henry et al 1995) and loose groups (Mulchaey et al 1996a) tend to be 
spiral poor, and led to the suggestion that spiral-rich compact groups might 
not be physically dense systems at all. 

The most extensive X-ray study of compact groups to date is that of Ponman 
et al (1996). These authors combined pointed and survey-mode observations of a 
complete sample of 85 HCGs and detected diffuse emission in 22 groups. They 
conclude that, when the detection limits are considered, diffuse emission 
is present in at least 75\% of the systems. Significantly, they detected 
diffuse emission in several spiral-rich groups.  In these the surface 
brightness is lower and the X-ray emission has a lower characteristic 
temperature, as would be expected given the lower velocity dispersions of 
spiral-rich compact groups. The diffuse X-ray luminosity was found to correlate 
with temperature, velocity dispersion, and spiral fraction, but not with 
optical luminosity. The last result suggests that the gas is mostly primordial 
and not derived from the galaxies. The correlations with temperature 
and velocity dispersion appear to be consistent with a single relation for
clusters and groups (Figure 5).

%--------------------------  figure 5
\begin{figure*}
\centerline{\psfig{figure=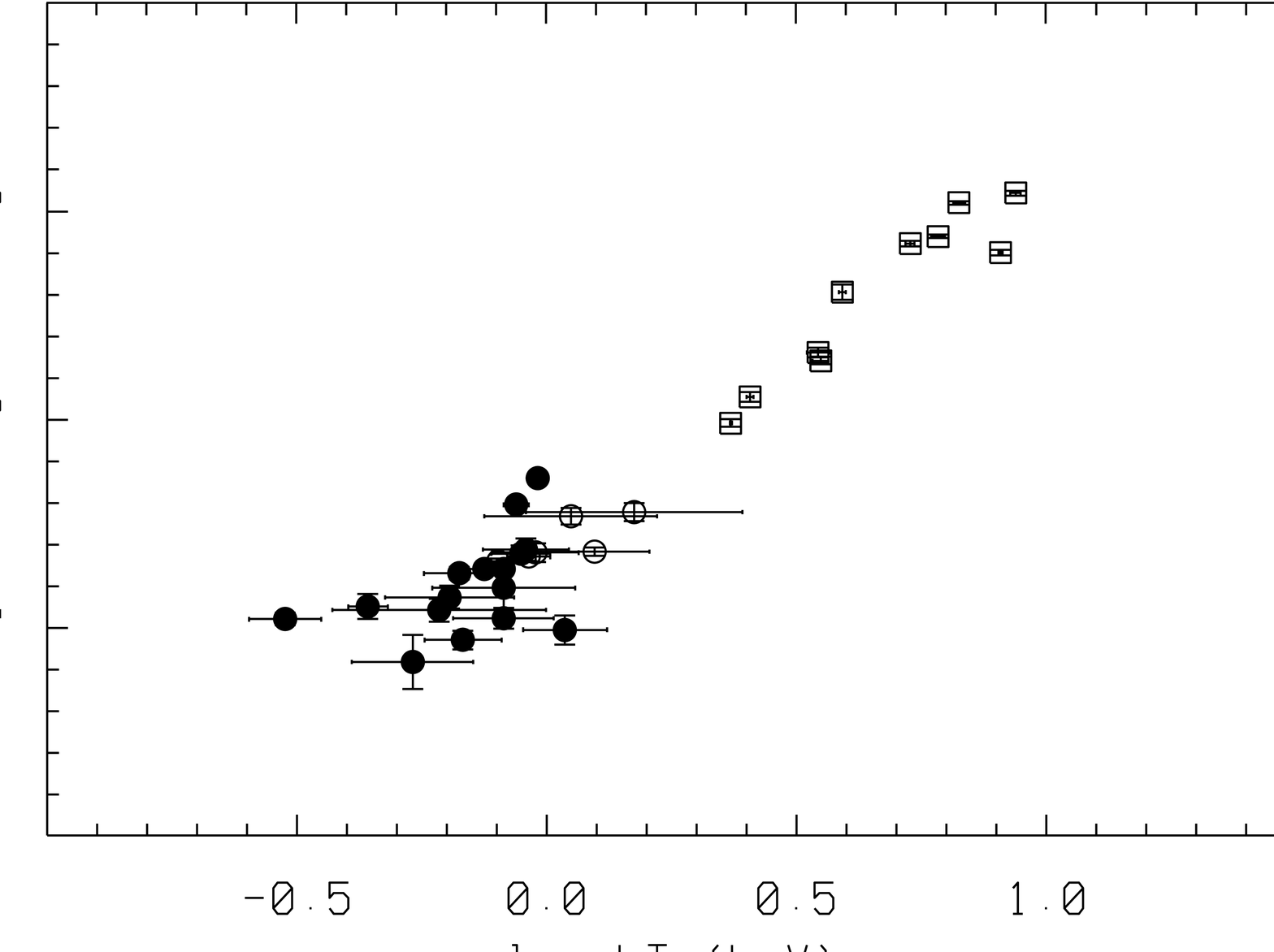,width=12cm,height=9cm}}
\vspace{1cm}
\caption{
Bolometric X-ray luminosity vs temperature. Filled circles indicate compact
groups, open circles indicate X-ray selected groups (Henry et al 1995) and
squares indicate clusters. X-ray data for the compact groups and clusters are 
taken from Ponman et al (1996). A single relation is consistent with clusters, 
groups and compact groups.
}
\end{figure*}
%---------------------------------

The total mass in compact groups typically exceeds the stellar and gas mass by 
an order of magnitude. Pildis et al (1995) derived baryon fractions of 12-19\%. 
Davis et al (1996) obtains 10-16\% for the NGC 2300 group. These are comparable 
to the fractions found for poor clusters (Dell'Antonio et al (1995) and are 
about half the typical values found for rich clusters. However, the derived 
baryon fraction depends sensitively on the radius within which it is measured 
and on the assumed background level (Henriksen \& Mamon 1994). Both the total 
mass, which is dominated by dark matter, and the gas mass continue to increase 
with radius. Consequently, both the baryon fraction and the gas fraction are
poorly determined. 

The contribution of compact groups to the X-ray luminosity function has been 
estimated by PBEB, who find that on the order of 4\% of the total luminosity in 
the range $10^{41} - 10^{43}$ erg s$^{-1}$ comes from HCGs. This is higher than 
the contribution of HCG galaxies to the local optical luminosity density 
estimated at 0.8\% by Mendes de Oliveira \& Hickson (1991), but it is 
comparable to the value found by Ribeiro et al (1994).

The metallicity inferred for the X-ray-emitting gas in compact groups is 
relatively low. PBEB obtain a mean metallicity of 0.18 solar, compared to the 
value, 0.3-0.4 solar, found in rich clusters. This is comparable to the low 
value ($< 0.11$ solar) found for the NGC 2300 group (Davis et al 1996). 
These figures suggest that the gas is largely primordial, a result
supported by the absence of a correlation between X-ray and optical
luminosity. However, given the limited spectral resolution of ROSAT, 
these low metalicities cannot yet be considered secure. The higher spectral 
resolution and sensitivity of the ASCA satellite should provide more
definitive results. Recent observations of HCG 51 and the 
NGC 5044 group found metal abundances comparable to those of clusters
(Fukazawa et al 1996). 

\section{PHYSICAL NATURE}

\subsection{Discordant redshifts}

The nature of the discordant redshift members of compact groups has been a 
subject of debate for many years (eg. Burbidge \& Burbidge 1961a,  Burbidge \& 
Sargent 1971, Nottale \& Moles 1978, Sulentic 1983). If the frequency of 
discordant galaxies is inconsistent with the statistics of chance projection, 
it might signify the need for new physical theories (Arp 1987), or for 
gravitational amplification of background galaxies (Hammer \& Nottale 1986). 
Initial estimates of the chance probability of finding discordant galaxies in 
groups like Stephan's Quintet, Seyfert's Sextet and VV 172, were very small 
(Burbidge \& Sargent 1971). However, such probabilities were recognized to be
difficult to determine reliably, because the a priori probability of {\it any}
particular configuration of galaxies is also very small (Burbidge \& Sargent 
1971). Only with a well-defined sample of groups and a complete characterization of selection effects, 
can meaningful estimates of the probabilities be made. The explicit selection 
criteria of the HCG catalog in principal makes this sample suitable for a
quantitative statistical investigation of the discordant-redshift question. 
Sulentic (1987) first concluded that the number of discordant redshifts in the 
catalog is too large to explain by chance. On the other hand, Hickson et al 
1988a and Mendes de Oliveira (1995), applying the selection criteria more 
rigorously, found no strong statistical evidence for this. Their result, 
however, may be biased by incompleteness in the HCG catalog: There seem to be 
too-few low surface brightness groups in the catalog, and the ``missing'' 
groups may have a higher fraction of discordant redshifts (Sulentic 1997).

In order to address the incompleteness issue, Iovino \& Hickson (1996) 
combined observational results from both the HCG and SCG catalogs with 
Monte-Carlo simulations. Their technique exploits the unbiased nature of the 
SCG catalog and the complete redshift coverage of the HCG sample. They conclude 
that for all except the two highest-surface-brightness quintets (Stephan's 
Quintet and Seyfert's Sextet), the number of discordant redshifts is consistent 
with chance projections. For these two, the chance probabilities are low. 
However, for both of these systems there is independent physical evidence that 
the discordant galaxies are at the cosmological distances that correspond to 
their redshifts and are therefore not group members (Kent 1981, Wu et al 1994).

One should not assume that the situation is now completely settled. Further 
studies will be possible when redshifts have been obtained for the SCG 
galaxies. There are still other questions that have not been adequately 
addressed, such as reported redshift quantization (Cocke \& Tifft 1983). 
However, at this point it appears that the frequency of discordant galaxies 
does not require a new interpretation of galaxy redshifts. In fact, physical 
evidence suggests the opposite. The discordant galaxies all have physical 
properties consistent with a cosmological distance. For example those with 
higher redshift tend to be smaller and fainter than other members of the group, 
and vice versa (Mendes de Oliveira 1995).

\subsection{Physical association and density}

Because we can measure only three phase-space dimensions for galaxies in 
compact groups (two components of position and one of velocity), the groups 
are subject to projection effects. Because of this, they may not be physically 
dense, or even physically related systems. 

The following interpretations have so far been suggested for compact groups:
\begin{enumerate}
\item{transient dense configurations (Rose 1977)}
\item{isolated bound dense configurations (Sulentic 1987, Hickson \& Rood 
1988)}
\item{chance alignments in loose groups (Mamon 1986, Walke \& Mamon 1989,
Mamon 1995)}
\item{filaments seen end-on (Hernquist et al 1995)}
\item{bound dense configurations within loose groups (Diaferio et al 1994,
Governato et al 1996)}
\end{enumerate}

Evidence for and against physical association and high density in the HCG 
sample, to 1988, was summarized by Hickson \& Rood (1988), and by Walke and 
Mamon (1989) respectively. Since that time, several new results have emerged.
From an analysis of optical images, Mendes de Oliveira \& Hickson (1994)
concluded that 43\% of all HCG galaxies show morphological features indicative 
of interaction and/or merging, and that 32\% of all HCGs contain three or more 
interacting galaxies. These percentages are likely to rise with more-detailed
studies and sophisticated image analysis (Longo et al 1994). This high 
frequency of interactions observed in compact groups is difficult to reconcile 
with the chance alignment and filament hypotheses, even if the alignments 
contain physical binaries (Mamon 1995). 

The high fraction of HCGs showing diffuse X-ray emission is very strong 
evidence that a large fraction of these systems are physically dense, and are 
not transient configurations or projection effects. Although the exact numbers 
are not final, due to the faintness of the sources and the problems of 
contamination by sources associated with the individual galaxies, it seems 
evident that many groups are dense bound systems. The correlations seen 
between X-ray and optical properties, and the fact that the X-ray properties
of compact groups are not inconsistent with those of clusters reinforces this 
conclusion.

Ostriker et al (1995) have argued that the relatively low X-ray luminosities of 
compact groups might not be due to a low gas fraction, but instead could be 
understood if the groups are filaments seen in projection (Hernquist et al 
1995). However, Ponman et al (1996) point out that in order to explain even 
the fainter compact groups, gas temperatures $T \sim 1$ keV and densities 
$n \sim 10^{-4}$ cm$^{-3}$ would be required. These appear to be ruled out by 
both observations (Briel \& Henry 1995) and simulations (Diaferio et al 1995,
Pildis et al 1996).

Even if compact groups are physically dense, they may not be as dense as they 
appear. As mentioned in Section 2, a sample of groups selected on the basis of 
high apparent density will be biased by the inclusion of looser systems which 
appear more compact due to geometrical or kinematic effects. Is this bias 
large? Its magnitude can be estimated as follows: Consider $n$ galaxies 
randomly located within a circle of radius $R$ on the sky. What is the 
probability $f(x,n)$ that they will fall within some circular subarea of radius 
$xR$? The answer can be obtained using analytic expressions derived by 
Walke \& Mamon (1989). From their Equations 1 and 6 (setting $N = n$ and 
${\cal N}_{ext} = 1$) we obtain
\be
  f(x,n) = {n!\over \pi^n} \int_0^x dr \int_0^{1-r} d\rho {d{\cal N}\over
    dr d\rho}
\ee
where
\be
  d{\cal N} = {2\pi^{n-1} n r^{2n-3}\over (n-2)!} 2\pi\rho d\rho dr
\ee
is the number of possible configurations with radius between $r$ and $r+dr$ 
and distance from the center between $\rho$ and $\rho + d\rho$. Here we have 
neglected a small edge contribution that is unimportant for small values of 
$x$ (Walke and Mamon's case 3). This gives
\be
  f(x,n) = n^2 x^{2n-2} \Bigr[ 1-4{n-1\over {2n-1}} x + {n-1\over n} x^2
  \Bigl]
\ee
Now, an observer would infer a galaxy space density that is higher by a factor 
$\beta = x^{-3}$, so the average apparent space density enhancement is
\ba
  <\beta> & = & \int_0^1 \beta {df(x,n)\over dx} dx \nonumber \\
    & = & \int_0^1 n^2 x^{2n-5} \Bigr[ 1-4{n-1\over {2n-1}} x + 
    {n-1\over n} x^2 \Bigl] dx \nonumber \\
    & = & {2n^2(n-1)\over (2n-5)(2n-3)(n-2)}
\ea
Thus we expect to typically overestimate the space density by about a factor 
of 12.0 for triplets (or quartets containing a physical binary), 3.2 for true 
quartets and 2.0 for quintets.

\section{COSMOLOGICAL IMPLICATIONS}

\subsection{Clustering and large-scale structure}

A key question that remains is the position of compact groups in the 
clustering hierarchy. Are compact groups distinct entities (Sulentic 1987) or 
an intermediate stage between loose groups and triplets, pairs and individual 
galaxies (Barnes 1989, White 1990, Cavaliere et al 1991, Rampazzo \& Sulentic 
1992, Diaferio et al 1994). Some compact groups are purely projection effects, 
others may be small clusters (Ebeling et al 1995), but most appear to be real. 
It seems that they can arise naturally from subcondensations in looser groups, 
but further studies are needed to better determine both the observed space 
density of groups as a function of population and the timescales involved in 
the evolutionary process.

This question is related to that of the formation mechanism of compact groups.
Two mechanisms have been discussed in the literature. Diaferio et al (1994) 
conclude that compact groups form continually from bound subsystems within 
loose groups. This gains some support from the observation (see Section 3) 
that most HCGs are embedded in loose groups, although it is not obvious that 
these loose groups are sufficiently rich (Sulentic 1997). Governato et al 
(1996), proposed a model in which merging activity in compact groups is 
accompanied by infall of galaxies from the environment. This naturally explains 
the observed mix of morphological types, and it allows compact groups to 
persist for longer times. 

Where do the Shakhbazian groups fit in this picture? Recent studies (Tikhonov 
1986, Amirkhanian \& Egikian 1987, Amirkhanian et al 1988, 1991, Amirkhanian 
1989,  Kodaira et al 1988-91, Stoll et al 1993-1996) show that these objects 
are typically compact clusters or groups of early-type galaxies. Although the 
systems were selected on the basis of red colors and compact appearance of 
their galaxies, both of these factors result from their large distances 
because K-corrections and contrast effects become significant. The galaxies 
are in fact relatively normal, although luminous (Del Olmo et al 1995). 
However, the number of blue (gas rich) galaxies in these systems does seem to 
be very small. Thus it appears the Shakhbazian groups are mostly small 
clusters, possibly intermediate in physical properties between classical 
compact groups and clusters.

\subsection{Galaxy evolution and merging}

If the groups are dynamically bound, galaxy mergers should commence within a 
few dynamical times (Carnevali et al 1981, Ishizawa et al 1983, Barnes 1985, 
Ishizawa 1986, Mamon 1987, 1990, Zheng et al 1993). Both N-body and 
hydrodynamic simulations indicate that the dark matter halos of individual 
galaxies merge first, creating a massive envelope within which the visible 
galaxies move (Barnes 1984, Bode et al 1993). Kinematic studies of loose groups 
(eg. Puche \& Carignan 1991) indicate that the dark matter is concentrated 
around the individual optical galaxies. In contrast, the X-ray observations 
indicate that in most compact groups, the gas and dark matter is more extended 
and is decoupled from the galaxies. This may explain the observation that 
galaxies in compact groups typically have mass-to-light ratios 30\% to 50\% 
lower than more isolated galaxies (Rubin et al 1991).

Is there any observational evidence that galaxies in compact groups are 
merging? By 1982 it was evident that first-ranked galaxies in compact
groups did not appear to be merger products, because the fraction of
first-ranked galaxies that are type E or S0 is the same as for the general
population of HCG galaxies (Hickson 1982). If mergers were a dominant effect, 
the first-ranked galaxies would be expected to be more often elliptical.
The same conclusion was reached by Geller \& Postman (1983) who found that the 
luminosities of first-ranked galaxies were consistent with a single luminosity
distribution for all group galaxies. Of course this may just mean that in small 
groups the first-ranked galaxy is not necessarily the most evolved. Rather, one 
should ask if {\it any} galaxies in compact groups show indications of merging. 
The relative paucity of merging galaxies in compact groups was first noted by 
Tikhonov (1987), from a visual inspection of optical images. Zepf \& Whitmore 
(1991) realized that elliptical galaxies formed by recent mergers of gas-rich 
systems should have bluer colors than normal. Examining the HCGs, they found 
only a small enhancement in the fraction of early-type galaxies having blue 
colors, a conclusion reinforced by an independent study by Moles et al (1994).  
On the other hand, Caon et al (1994) argued that the large effective radii of 
compact group elliptical galaxies is indicative of an origin by merging or 
accretion of companions.

Zepf (1993) estimated that roughly 7\% of the galaxies in compact groups are 
in the process of merging. This conclusion was based on roughly consistent 
frequencies of {\it (a)} optical signatures of merging, {\it (b)} warm 
far-infrared colors, and {\it (c)} sinusoidal rotation curves. However, few 
galaxies show all of these effects simultaneously. The merging fraction may 
thus be as high as 25\% if one allows that any one of these criteria would be
considered to be sufficient to indicate a merger (Hickson 1997). Given the 
small numbers of objects in these studies, it is fair to say that the fraction 
of merging galaxies is highly uncertain at present. It seems safe to conclude 
that current observations do not rule out a significant amount of merging in 
compact groups.

Detailed studies of individual compact groups can be quite revealing. Many 
galaxies that at first appear normal are revealed to have peculiar morphology 
or spectra when examined more closely. Many, perhaps most, compact groups 
clearly contain galaxies that are dynamically interacting. However, the groups 
likely span a range of evolutionary states. At the extreme end are high-density 
groups like Seyfert's Sextet, HCG 31, HCG 62, HCG 94 (Pildis 1995) and HCG 95 
(Rodrigue et al 1995) in which we find strong gravitational interactions. At 
the other end are lower density compact groups, such as HCG 44, which most 
likely are in a less-advanced stage of evolution. This picture is supported by 
radio observations: Seyfert's Sextet and HCG 31 are both embedded in extended 
HI clouds whereas in HCG 44 the HI is associated with individual galaxies 
(Williams et al 1991).

It seems clear that the groups as we now see them can persist for only a 
fraction of a Hubble time. Simulations indicate that merging should destroy the 
group on a time scale $t_m$ that is typically an order of magnitude larger 
than $t_d$, depending on the distribution of dark matter (Cavaliere et al 1983, 
Barnes 1984, Navarro et al 1987, Kodaira et al 1990) and initial conditions 
(Governato et al 1991). Assuming that the groups are in fact bound dynamical 
systems, we can draw two conclusions: {\it (a)} There must be an ongoing 
mechanism for forming or replacing compact groups, and {\it (b)} there must be 
a significant population of relics of merged groups.

What are the end-products of compact groups? It is tempting to identify
them with field elliptical galaxies, following a suggestion first made
by Toomre (1977). Simulations (Weil \& Hernquist 1994) indicate that multiple 
mergers in small groups of galaxies best reproduce the observed kinematical 
properties of elliptical galaxies. The resulting galaxies are predicted to 
possess small kinematic misalignments, which can be detected by detailed 
spectroscopic and photometric studies. Neverthess, it remains to be 
demonstrated that these merger remnants can reproduce the tight correlation 
between size, luminosity and velocity dispersion found in present-day 
elliptical galaxies.

If compact groups have lifetimes on the order of $t_m$, and form continuously, 
then the number of relics, per observed group, is expected to be on the order 
of $(H_0t_m)^{-1}$. Thus, the number of relics could exceed that of present day 
groups by as much as an order of magnitude. Mamon (1986) estimated that, if all
HCGs are real, then the relics would account for about 25\% of luminous field 
elliptical galaxies. As we have seen, the true space density of compact groups 
is uncertain by at least a factor of two, and may be underestimated because of
selection biases. There is then the potential problem of producing too many 
relics.

A second problem is the fact that the integrated luminosities of compact groups 
are typically a factor of three to four times greater than luminosities of 
isolated elliptical galaxies (Sulentic and Raba\c{c}a 1994). It is possible 
that interaction-induced star formation has boosted the luminosities of some 
compact-group galaxies, and that some degree of fading of the merger product 
is expected. However, at this point it is not clear whether or not the relics 
can be identified with isolated elliptical galaxies.

Despite these problems, a fossil compact group may have actually been found.
Ponman et al (1995) have detected a luminous isolated elliptical galaxy 
surrounded by diffuse X-ray emission which is consistent with the expected 
end-product of a compact group. If more objects like this are found, it may be 
possible to compare their space density with that expected for compact group 
relics.

\subsection{Role in galaxy formation and evolution}

Interactions are often implicated in the development of active nuclei in 
galaxies (eg. Freudling \& Almudena Prieto 1996). The HCG catalog includes 
several examples of compact groups containing both starburst galaxies and AGN. 
Several recent examples of associations between starburst galaxies or AGN and 
what appear to be compact groups have been reported: Del Olmo \& Moles (1991)
have found a broad-line AGN in Shakhbazian 278; Zou et al (1995) find that 
the luminous infrared source IRAS 23532 coincides with a compact group that 
includes a Seyfert 1 as well as a starburst galaxy. If this association extends to QSOs, one 
would expect to find numerous compact groups at redshifts $z \sim 2$, 
where the comoving number density of QSOs peaks (eg. Hartwick \& Schade
1984). The tendency for QSOs to have close companions has been known for some
time (eg. Stockton 1982, Bahcall et al 1997). Recently, several examples of 
compact groups associated with luminous infrared galaxies, AGN and QSOs at 
$z \simeq 2$ have been found using HST (Pascarelle et al 1996, Francis et al 
1996, Matthews et al 1994, Tsuboi \& Nakai 1994, Hutchings 1995, Hutchings 
et al 1995).

These observations provide support to the idea that tidally-triggered star 
formation is a predominant factor in the galaxy formation process 
(Lacey \& Silk 1991, Lacey et al 1993). In this model disk star formation 
occurs relatively late, after the compact group has formed and tidal 
interactions are strong. This seems at least qualitatively consistent with the
fragmentary nature of high-redshift galaxies observed with the Hubble Space 
Telescope (Schade et al 1995), although these fragments appear to be 
much less luminous and more irregular than most present-day compact group 
galaxies. The model also offers a possible explanation for the excess numbers 
of faint blue galaxies found in field galaxy count as dwarf galaxies undergoing 
star formation at a redshift of $z \simeq 1$.

Compact groups may possibly play a role in the formation of other systems. We 
have seen that giant galaxies may be formed as the end product of compact-group 
evolution. At the other end of the scale, dwarf galaxies have physical 
properties distinct from normal galaxies, which suggests a unique formation 
mechanism. One possibility is that they form during gravitational interactions
from tidal debris (Duc \& Mirabel 1994). If this is the case, one would expect 
to find evidence for this in compact groups of galaxies. From an examination of 
condensations in tidal tails, Hunsberger et al (1996) concluded that the 
fraction of dwarf galaxies produced within tidal debris in compact groups is 
not negligible. There is also evidence that star clusters form from tidal 
debris. Longo et al (1995) have found an excess population of unresolved blue
objects around HCG 90 which appear to be recently formed star clusters. These 
may be similar to the population of new star clusters recently reported in the 
merger remnant NGC 7252 (Whitmore et al 1993).

\subsection{Gravitational lensing}

Because compact groups have a high galaxy surface density, they may form
effective gravitational lenses. Gravitational amplification of background
field galaxies was proposed by Hammer and Nottale (1986) as a possible 
explanation for the presence of the high-redshift discordant member of this 
group. Mendes de Oliveira and Giraud (1994) and Montoya et al (1996) find that 
most HCGs are too nearby to produce strong lensing effects. However, because 
the critical mass density required for strong lensing depends reciprocally on 
distance, analogous systems 5-10 times more distant should produce a 
non-negligible  fraction of giant arcs.

\subsection{Cosmology}

Studies of small groups may provide clues to the overall structure of the 
universe. The baryon fractions found in clusters of galaxies appear to be 
inconsistent with a density parameter $\Omega = 1$, unless the dark matter is 
more prevalent outside clusters (White 1992, Babul \& Katz 1993). Compact 
groups provide a means to study dark matter in such regions. The baryon 
fractions found for compact groups do appear to be lower than those for 
clusters. David et al (1995) argued that the gas is the most extended component; 
galaxies being the most compact and the dark matter being intermediate. They 
concluded that the baryon fraction approaches 30\% on large enough scales, 
which is comparable to the values found for clusters. Given the constraints of 
standard Big-Bang nucleosynthesis this would imply that the density parameter 
$\Omega$ is at most 0.2. On the other hand, the infall picture of compact-group 
evolution (Governato et al 1996) requires a high-density $\Omega \sim 1$
universe. In a low-density universe the infall rate is insufficient. As there 
is at present no other clear mechanism for avoiding the overproduction of 
relics by merging compact groups, this may be a strong argument for a 
high-density universe.

During the last two decades we have seen a resurgence of interest in compact 
groups. While initially little more than a curiosity, these systems are now 
viewed as potentially important sites of dynamical evolution, shaping the 
structure of many galaxies. It now seems clear that while many compact groups 
are contaminated by projections, a large fraction of at least the 
high-surface-brightness HCGs are physically dense.  They form by gravitational 
relaxation processes within looser associations of galaxies. The densest are 
generally in an advanced stage of evolution characterized by strong 
interactions, starburst and AGN activity, stripping of stellar and dark matter 
halos, and merging. They contain large amounts of dark matter and primordial 
X-ray-emitting gas trapped within the gravitational potential well.

Despite this progress, many questions remain unanswered. What are the end
products of compact group evolution, and do they have properties consistent
with any know population of objects? What is the space density of such relics?
Where do compact groups fit in the overall clustering hierarchy? What is
their role in the evolution of galaxies both past and present? Given the
current interest and research activity in this area, it is likely that
many of these questions may soon be addressed.

\begin{acknowledgements}

It is a pleasure to thank the Observatories of Brera and Capodimonte for 
hospitality during the initial work on this review. I have benefitted from 
discussions with many individuals, but I would like to acknowledge particularly 
the contributions of A Iovino, E Kindl, G Longo, G Mamon, C Mendes de Oliveira,
TK Menon, G Palumbo, H Rood, and J Sulentic. I thank G Mamon, A Sandage
and J Sulentic for providing helpful comments on an earlier version of the 
manuscript. Financial support was provided by the Natural Sciences and 
Engineering Research Council of Canada and NATO.

\end{acknowledgements}

\end{document}